\documentclass[11pt]{article}
\usepackage{times}
\usepackage{pdfpages}
\usepackage{setspace}
\usepackage{sectsty}
\usepackage{enumitem}

\sectionfont{\fontsize{11}{15}\selectfont}
\subsectionfont{\fontsize{11}{15}\selectfont}

\usepackage{wrapfig}
\usepackage{lipsum}
\usepackage{graphicx}

\usepackage{array}
\usepackage{subfigure}
\usepackage{booktabs}
\usepackage{multicol}
\usepackage{multirow}
\usepackage{threeparttable}
\usepackage[small, bf]{caption}
\usepackage[tableposition=top]{caption}
\usepackage{booktabs}
\usepackage{siunitx}

\usepackage{pgfgantt}

\usepackage{wrapfig}
\usepackage{times}
\usepackage{url}
\usepackage{soul}
\usepackage{color}
\begingroup
\makeatletter
\g@addto@macro{\UrlSpecials}{%
  \endlinechar=13 \catcode\endlinechar=12
  \do\%{\Url@percent}\do\^^M{\break}}
 \gdef\Url@percent{\@ifnextchar^^M{\@gobble}{\mathbin{\mathchar`\%}}}%
\endgroup %
\usepackage{array}
\usepackage{subfigure}

\usepackage{array}
\newcolumntype{L}[1]{>{\raggedright\let\newline\\\arraybackslash\hspace{0pt}}m{#1}}

\usepackage[headheight=14pt,tmargin=1in,bmargin=1in,lmargin=1in,rmargin=1in]{geometry}

\usepackage{fancyhdr}
\usepackage{datetime}
\usepackage{indentfirst}
\usepackage[sort&compress, square, comma, numbers]{natbib}

\usepackage[compact]{titlesec}

\usepackage{titlesec}
\titlespacing*{\section}{0pt}{0.1\baselineskip}{0.1\baselineskip}
\titlespacing*{\subsection}{0pt}{0.2\baselineskip}{0.1\baselineskip}
\titlespacing*{\subsubsection}{0pt}{0.2\baselineskip}{0.1\baselineskip}
\titlespacing*{\paragraph}{0pt}{0.2\baselineskip}{0.2\baselineskip}

\usepackage{blindtext}

\usepackage{threeparttable}
\usepackage{soul}
\usepackage{booktabs}
\usepackage{multirow}
\usepackage{caption}
\usepackage{listings}
\usepackage[hidelinks]{hyperref}
\makeatletter
\g@addto@macro{\UrlBreaks}{\UrlOrds}
\makeatother

\begin{document}

\thispagestyle{empty}
\begin{center}
{\bf \Large Did They Really Tweet That? \\ Querying Fact-Checking Sites and Politwoops to Determine Tweet Misattribution}\\

\vspace{5mm}
Caleb Bradford, cbrad022@odu.edu\\
Department of Computer Science, Old Dominion University\\
\hfill \break
Michael L. Nelson, mln@cs.odu.edu\\
Department of Computer Science, Old Dominion University\\
\end{center}

\begin{abstract}
Screenshots of social media posts have become common place on social media sites. While screenshots definitely serve a purpose, their ubiquity enables the spread of fabricated screenshots of posts that were never actually made, thereby proliferating misattribution disinformation. With the motivation of detecting this type of disinformation, we researched developing methods of querying the Web for evidence of a tweet's existence. We developed software that automatically makes search queries utilizing the body of alleged tweets to a variety of services (Google, Snopes built-in search, and Reuters built-in search) in an effort to find fact-check articles and other evidence of supposedly made tweets. We also developed tools to automatically search the site Politwoops for a particular tweet that may have been made and deleted by an elected official. In addition, we developed software to scrape fact-check articles from the sites Reuters.com and Snopes.com in order to derive a ``truth rating" from any given article from these sites. For evaluation, we began the construction of a ground truth dataset of tweets with known evidence (currently only Snopes fact-check articles) on the live Web, and we gathered MRR and P@1 values based on queries made using only the bodies of those tweets. These queries showed that the Snopes built-in search was effective at finding appropriate articles about half of the time with $MRR=0.5500$ and $P@1=0.5333$, while Google when used with the site:snopes.com operator was generally effective at finding the articles in question, with $MRR=0.8667$ and $P@1=0.8667$.

\end{abstract}
\setcounter{page}{1}
\section{Introduction}

Screenshots of other social media posts are common in social media. This is a practice born out of necessity. Since social media companies are not incentivized to provide methods of sharing posts between different sites, they do not provide native functionality to cross-share posts. However, nearly every social media site provides a feature to share images. Thus, screenshots have provided users with a way to bypass social media's lack of interoperability (Figure \ref{fig:bidenreshare}). Screenshots also act as an accessible way to ``archive" posts that eventually became unavailable on the live Web. This can occur if the original poster deleted their post, or if the account of the original poster was removed from the platform entirely (Figure \ref{fig:deletedexample}). Screenshots even have their use in sharing posts that technically could be shared using a native feature of the site. Sharing tweets on Twitter could be done via Twitter's ``quote tweet" (Figure \ref{fig:qtexample}), ``retweet", and reply functionalities. However, this provides the original poster with engagement that a user sharing their post may want to avoid giving them. Thus, simply sharing a screenshot of their tweet instead accomplishes practically the same task without propagating it. \\

\begin{center}
\includegraphics[scale=0.5]{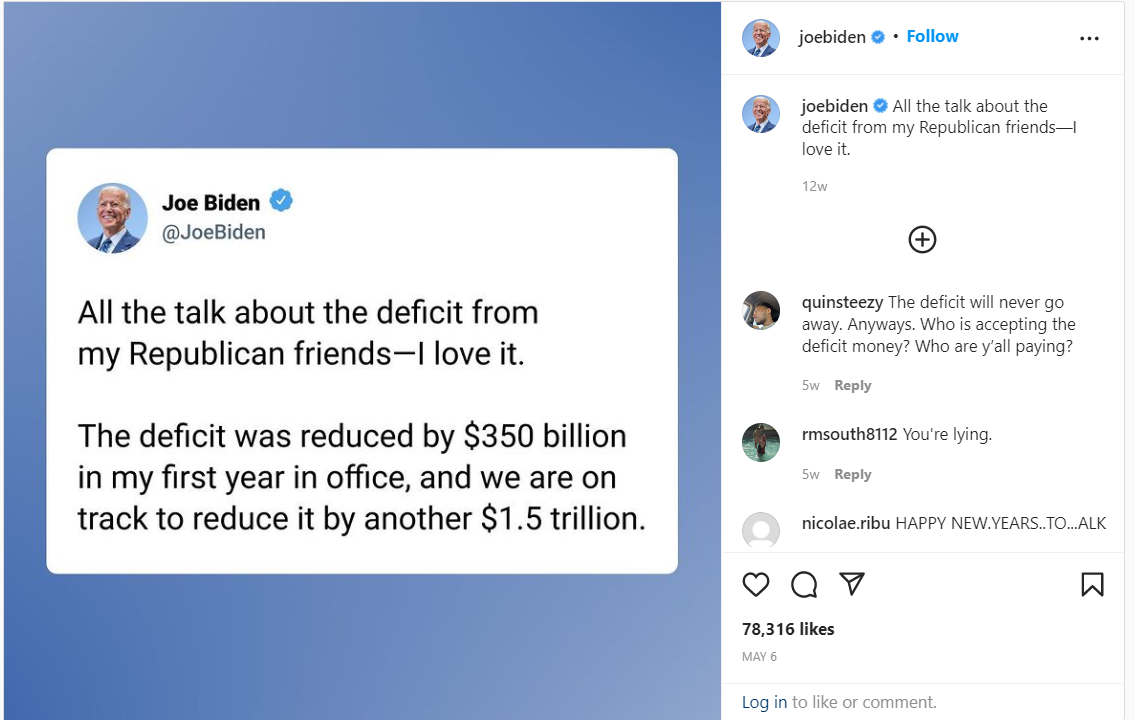}%
\captionof{figure}{A Joe Biden tweet being reshared as a screenshot by Joe Biden on Instagram.\protect\footnotemark}\label{labelname}%
\label{fig:bidenreshare}
\end{center}
\footnotetext{\url{https://www.instagram.com/p/CdPFNhHlQig/}}

\begin{center}
\includegraphics[scale=0.5]{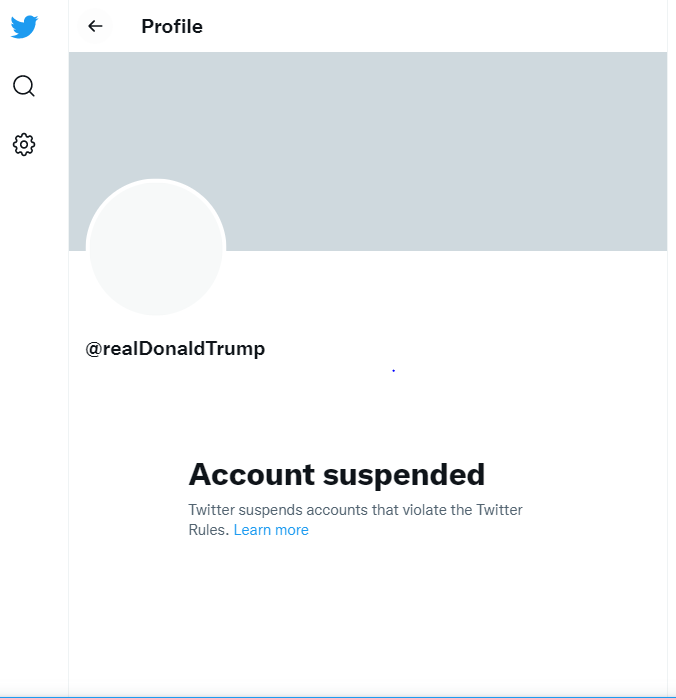}%
\captionof{figure}{Donald Trump's tweets are no longer available on the live Web since his account was suspended.\protect\footnotemark}\label{labelname}%
\label{fig:deletedexample}
\end{center}
\footnotetext{\url{https://twitter.com/realdonaldtrump}; see \cite{twitter-suspends-trump, garg2021replaying, kriesberg2022second}}

\begin{center}
\includegraphics[scale=1]{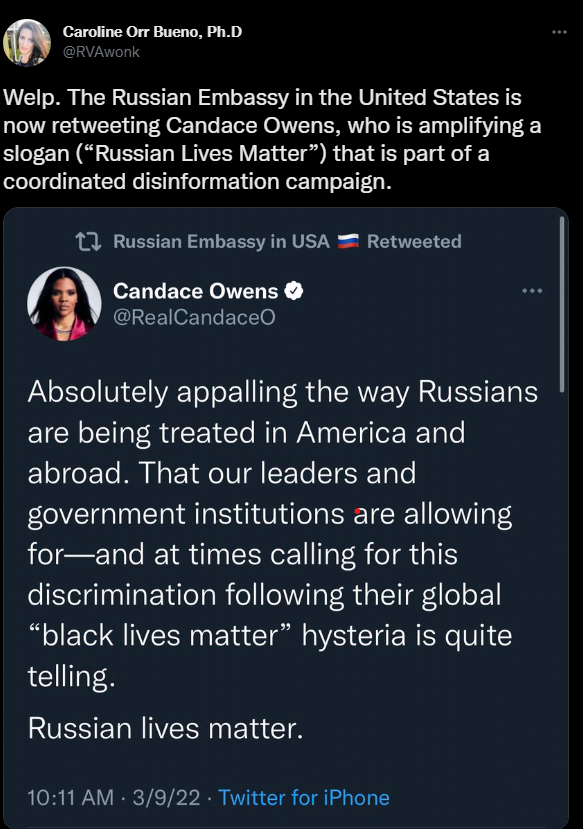}%
\captionof{figure}{A Twitter user sharing a screenshot of a tweet to avoid providing engagement to the original poster.\protect\footnotemark}\label{labelname}%
\label{fig:qtexample}
\end{center}
\footnotetext{\url{https://twitter.com/rvawonk/status/1503227687917305863}}

Sharing screenshots of social media posts has been established as a common practice for users of any social media site. However, screenshots of social media posts cannot be considered to be concrete evidence of that post being made because these screenshots can be fabricated. Fabricated screenshots are created for a variety of reasons, one being satire. Figure \ref{fig:satireexample} shows a fake Donald Trump tweet meant to make light of the Suez Canal obstruction that occurred in 2021.\footnote{\url{https://en.wikipedia.org/wiki/2021_Suez_Canal_obstruction}} 

The prevalence of both real and fake screenshots on social media has facilitated the rise of a unique type of disinformation: misattribution disinformation by means of fake screenshots tricking social media users into believing the false attribution present in the screenshot. A prominent example of this occurred when a screenshot of a fake post by Ted Cruz. Figure \ref{fig:tedcruz} was shared in response to his vacation to Mexico while Texas was going through a winter storm that disrupted much of its infrastructure.\footnote{\url{https://en.wikipedia.org/wiki/2021_Texas_power_crisis}} This fake screenshot gained massive exposure when it was tweeted by the account @rezaaslan with text falsely stating that it was a real tweet (Figure \ref{fig:rezaaslan}). This tweet incited mass engagement on Twitter, receiving over 100,000 likes before it was eventually deleted. However, the damage had already been done. Tens of thousands of Twitter users saw this tweet before its deletion, deceiving many of them into believing that the screenshot was authentic.

\begin{center}
\includegraphics[scale=0.5]{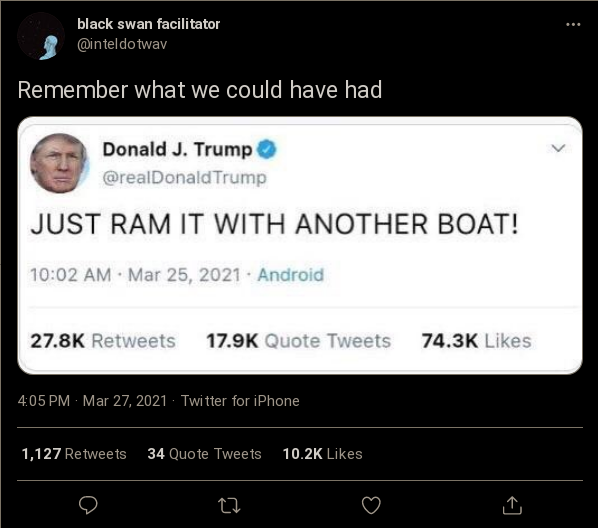}%
\captionof{figure}{A Twitter user sharing a comedic fake screenshot}\label{labelname}%
\label{fig:satireexample}
\end{center}

\begin{center}
\includegraphics[scale=0.5]{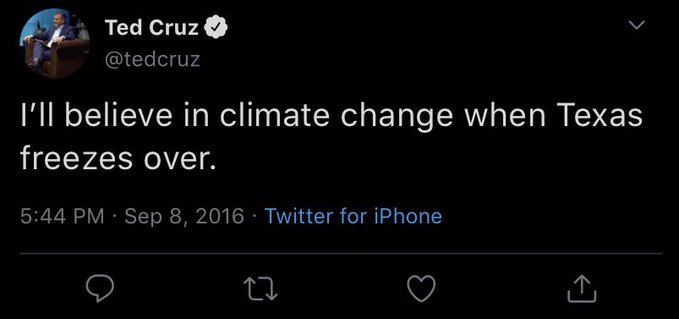}%
\captionof{figure}{A screenshot depicting a Ted Cruz tweet that was never made}\label{labelname}%
\label{fig:tedcruz}
\end{center}

\begin{center}
\includegraphics[scale=0.5]{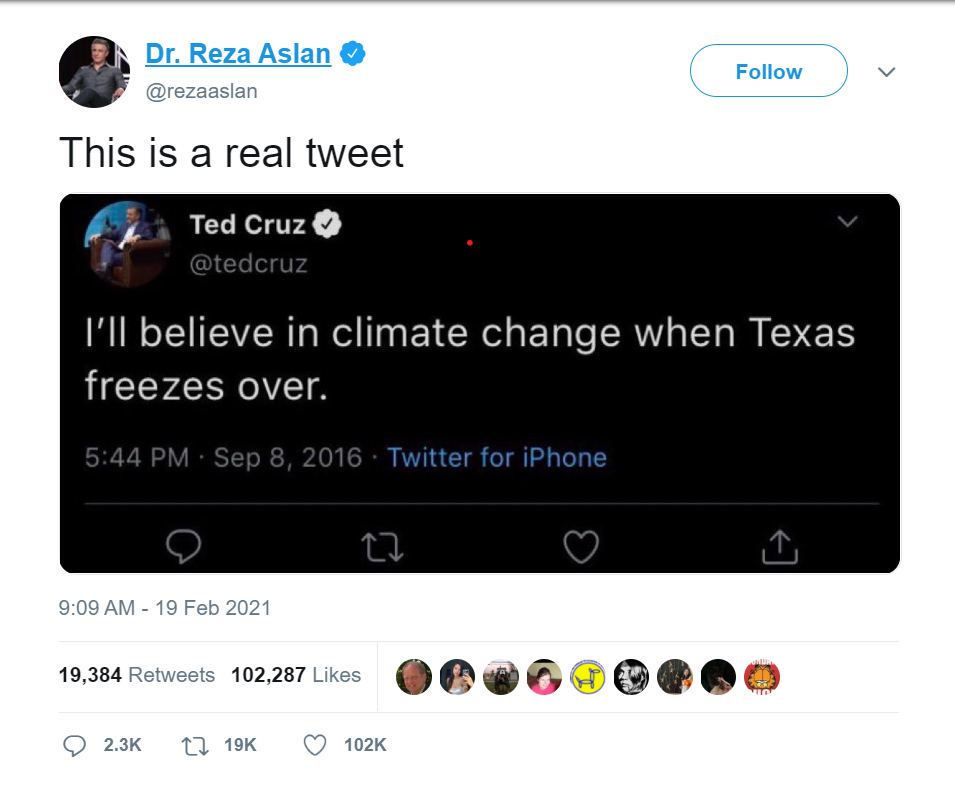}%
\captionof{figure}{@rezaaslan sharing the fake Ted Cruz tweet, inciting massive engagement.\protect\footnotemark}\label{labelname}%
\label{fig:rezaaslan}
\end{center}
\footnotetext{\url{https://web.archive.org/web/20210220030114/https://twitter.com/rezaaslan/status/1362811514172755969}}

This is a particularly dangerous, yet subtle form of disinformation because well-made fake screenshots are visually indistinguishable from a screenshot of a real post. Creating these screenshots is not even particularly difficult. This is especially true for posts on Twitter since tweets are intentionally designed to be simple posts that are easily digestible. In fact, fabricating convincing screenshots of social media posts is not a difficult task thanks to a variety of accessible Web applications such as Tweetgen (Figure \ref{fig:tweetgen}), which can generate fake Twitter posts, and Generate Status (Figure \ref{fig:generatestatus}, \ref{fig:generatestatusfb}), which can generate a wide variety of fake social media posts.\\

\begin{center}
\includegraphics[scale=0.5]{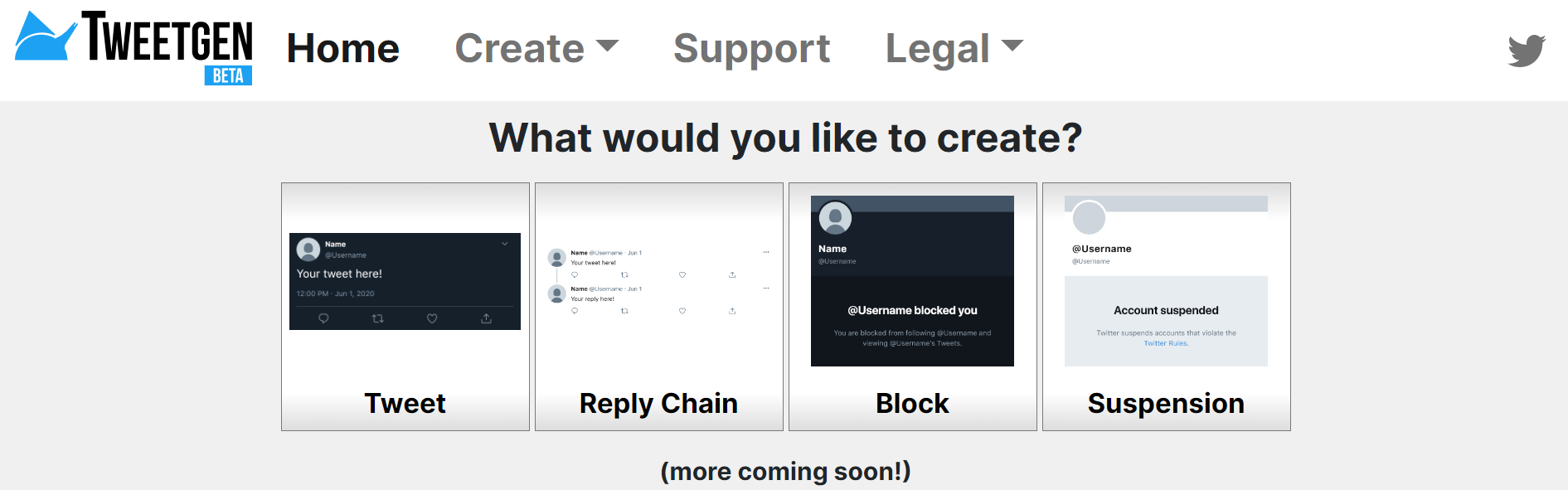}%
\captionof{figure}{Tweetgen can generate fabricated tweets.\protect\footnotemark}\label{labelname}%
\label{fig:tweetgen}
\end{center}
\footnotetext{\url{https://www.tweetgen.com/}}

\begin{center}
\includegraphics[scale=0.5]{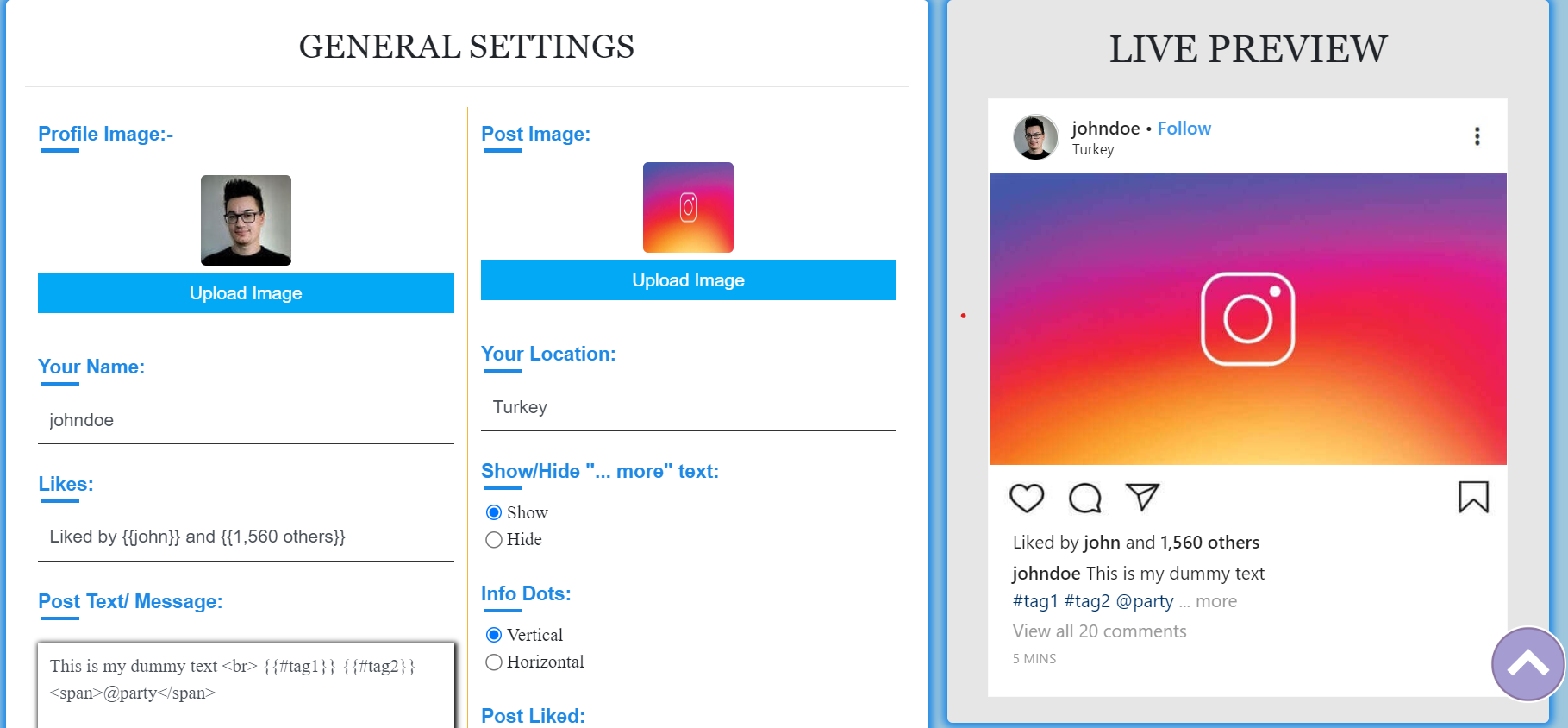}%
\captionof{figure}{Generate Status can fabricate Instagram posts.\protect\footnotemark}\label{labelname}%
\label{fig:generatestatus}
\end{center}
\footnotetext{\url{https://generatestatus.com/generate-fake-instagram-post/}}

\begin{center}
\includegraphics[scale=0.25]{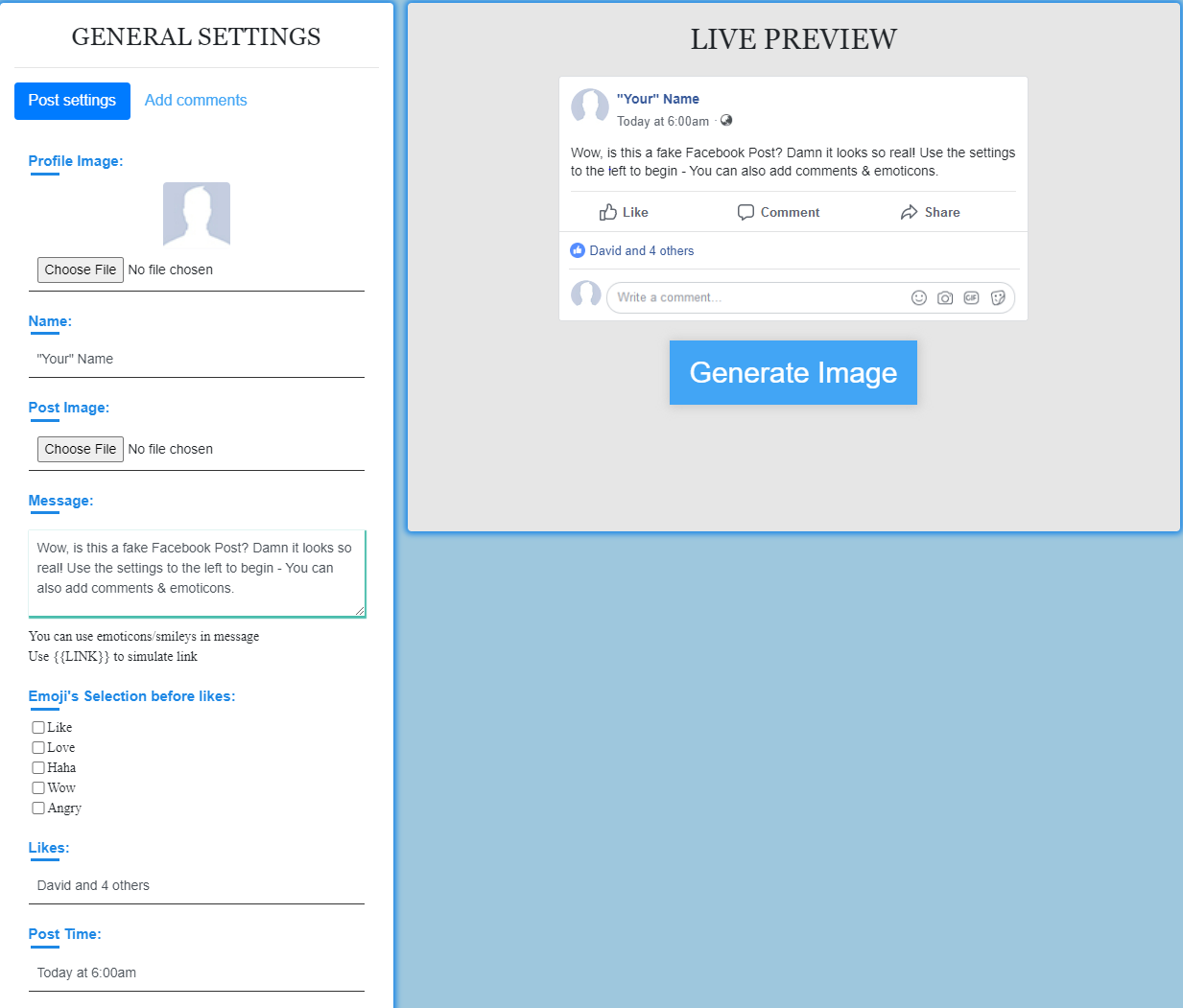}%
\captionof{figure}{Generate Status can fabricate Facebook posts.\protect\footnotemark}\label{labelname}%
\label{fig:generatestatusfb}
\end{center}
\footnotetext{\url{https://generatestatus.com/fake-facebook-post-generator/}}

The spread of this novel type of disinformation motivates the creation of a system that is capable of determining the veracity of a post being depicted in a screenshot given only that screenshot. We will refer to this system as \emph{SSAuth} from here on. The algorithm for \emph{SSAuth} involves first classifying the social media platform that the screenshot was supposedly taken from, then extracting all relevant metadata from the screenshot, and finally using that metadata to make a determination about the authenticity of the alleged post. Our research to date has focused on completing the final step of verifying the veracity of posts on Twitter, although other social media platforms are part of our future plans. 

One methodology for handling the verification of Twitter posts involves the fact that posts made by social media accounts run by well-known figures, including celebrities and political officials, often attract media attention. This is true of both authentic posts and fabricated posts that were widely circulated. This coverage results in content on the Web that explicitly states whether or not a post in question was actually made. News organizations often produce fact-check articles solely meant to verify attribution of supposedly made social media posts. News sites that often produce this type of content include Snopes.com\footnote{\url{https://snopes.com/}} and Reuters.com.\footnote{\url{https://reuters.com/}} There are also other sites on the Web that confirm the authenticity of social media posts. Existence of a tweet on Web archives such as the Internet Archive\footnote{\url{https://web.archive.org/}} and archive.today\footnote{\url{https://archive.ph/}} directly implies that that tweet was made. The same is true of tweets that exist on the deleted tweet tracker for politicians called Politwoops.\footnote{\url{https://projects.propublica.org/politwoops/}}

The existence of these sites that have content specifically focused on verifying social media post attribution motivates developing methods of leveraging that content in \emph{SSAuth}. Our research has focused on the construction and automation of search queries meant to find this evidence, as well as the parsing of this evidence for a ``truth value" in order to classify a tweet's veracity. 

\section{Background}
\subsection{The focus of this research}
Here, we define the critical distinction of the type of disinformation/misinformation that we are working to detect. This research is not concerned with verifying the actual validity of claims made within a screenshot of a supposed social media post. We are specifically concerned with verifying the attribution present in a screenshot of a social media post. In other words, we want to prove that this screenshot depicts a social media post that was actually made by the account present in the screenshot. In the example of the fake Ted Cruz tweet (Figure \ref{fig:tedcruz}), this is the distinction between determining the validity of the statement "I'll believe in climate change when Texas freezes over," and determining whether or not Ted Cruz actually tweeted that statement. The latter is what we are interested in verifying for this project. 

\subsection{Choosing fact-check articles for our purposes}
Fact-check articles are articles generated by news organizations that are specifically meant to spread awareness about the validity of a certain claim. These articles often consist of a statement of the claim in question, their decided ``truth rating" of the claim, and an explanation of the work done to come to those conclusions. This research is concerned with fact-check articles that verify tweet attribution, which are often generated in response to the spread of more outrageous tweets supposedly made by well-known accounts

Our methodology of searching the Web for evidence of tweet attribution necessitates choosing sites that we can accept evidence from. There are a few criteria that we considered in our selection.

In our search for traditional fact-checking sites, we wanted to ensure that these sites actually produced content that concerned tweet attribution. The distinction did need to be made between content that fact-checks the body of a tweet and content that verifies whether or not a tweet was actually made. However, it is worth noting that the existence of an article fact-checking the content of a tweet usually implied that the tweet in question was actually made, thereby necessitating an article explaining the validity of its content. 

It would also be advantageous for us to choose fact-checking sites that have a method of explicitly classifying their truth ratings within an article so that they can consistently be programmatically extracted. This simplifies the process of scraping a truth rating given any fact-check article from a particular site. Future work for this project may include development of more nuanced approaches to the rating extraction process that would involve deriving a truth rating from any generic fact-check article, but that remains outside the scope of the work done thus far.

We developed tools for two traditional fact-checking sites so far, that being Snopes.com (Figure \ref{fig:snopesarticle}) and Reuters.com (Figure \ref{fig:reutersarticle}). Both of these sites provide sizeable volumes of articles concerning tweet attribution and both also provide explicit parts of their articles in which they state the ``truth rating" of the claim being discussed in the article. 

\begin{center}
\includegraphics[scale=0.5]{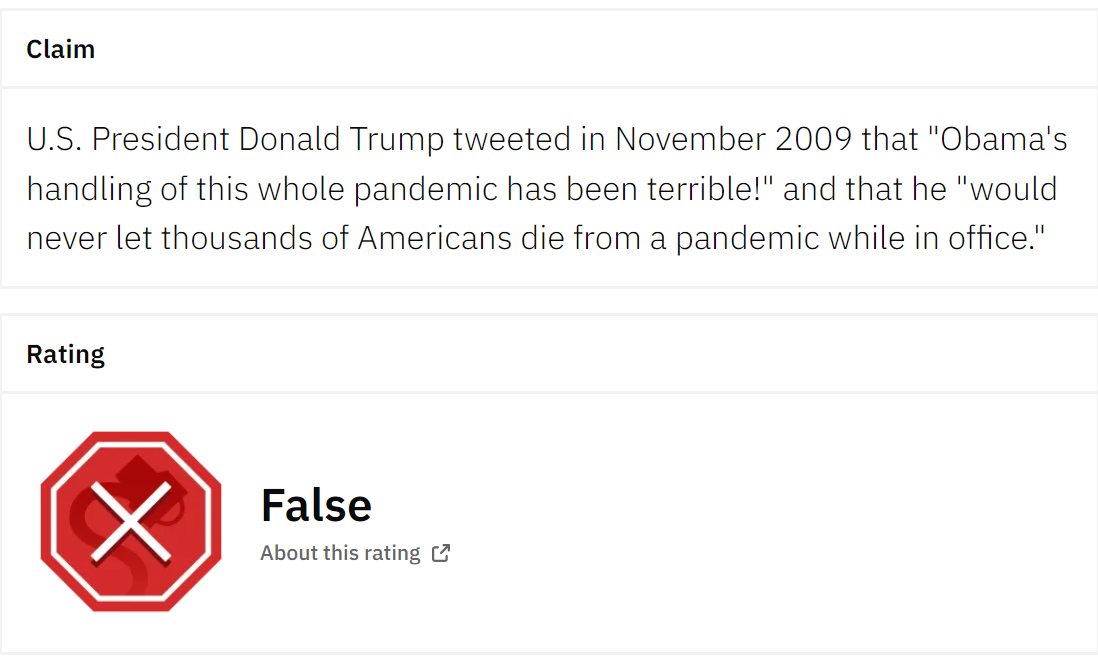}%
\captionof{figure}{A fact-check from Snopes.com\protect\footnotemark}\label{labelname}%
\label{fig:snopesarticle}
\end{center}
\footnotetext{\url{https://www.snopes.com/fact-check/2009-trump-tweet-pandemic/}}

\begin{center}
\includegraphics[scale=0.5]{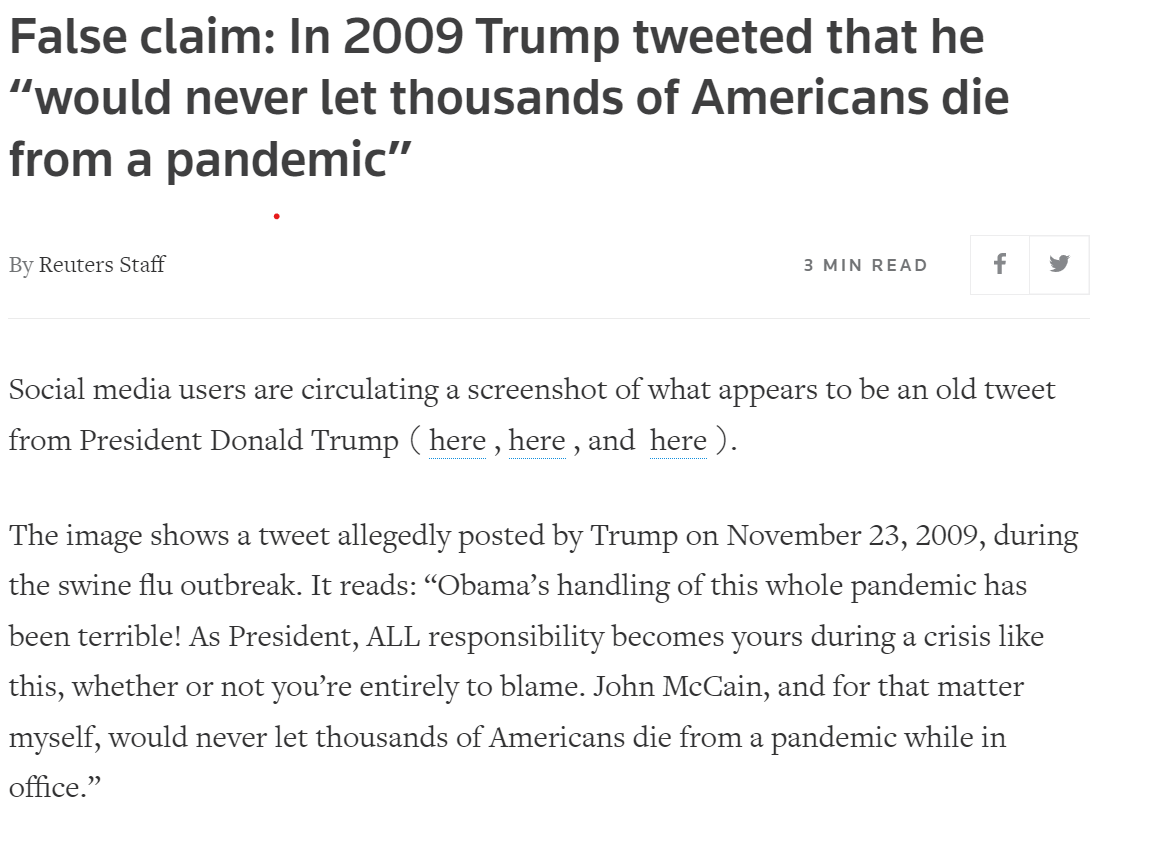}%
\captionof{figure}{A fact-check from Reuters.com\protect\footnotemark}\label{labelname}%
\label{fig:reutersarticle}
\end{center}
\footnotetext{\url{https://www.reuters.com/article/uk-factcheck-trump-tweet-thousands-die/false-claim-in-2009-trump-tweeted-that-he-would-never-let-thousands-of-americans-die-from-a-pandemic-idUSKCN2242AK}}

\subsection{Deleted Tweets at Politwoops}
In addition to fact-checking sites, there are other sites that contain evidence that we can leverage to prove or disprove tweet attribution. One of the most potentially fruitful of these sites for our uses is a project created by Probublica called Politwoops (Figure \ref{fig:politwoops}). Politwoops crawls the Twitter accounts of elected officials and candidates for office in order to detect when they delete a tweet. If it detects that a tweet made by an account that they are tracking is deleted, then they will preserve that tweet in order to hold political officials accountable for what they post on Twitter. Politwoops is especially useful for our purposes because existence of a tweet on Politwoops directly implies that the tweet was not only made, but it was intentionally deleted by the original poster, potentially because they did not want those words associated with them. For this reason, we developed tools to search for a specific tweet on Politwoops. 

\begin{center}
\includegraphics[scale=0.5]{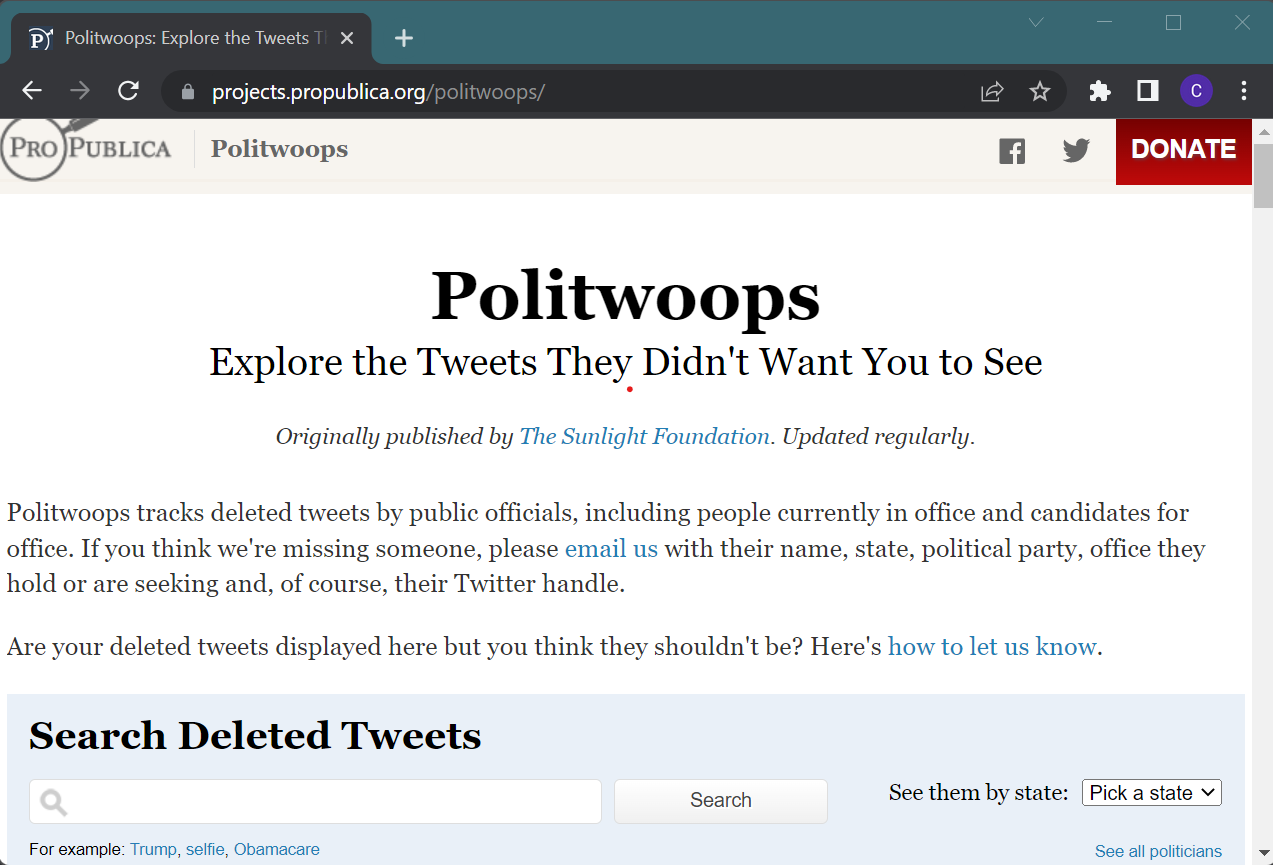}%
\captionof{figure}{The homepage of Politwoops.\protect\footnotemark}\label{labelname}%
\label{fig:politwoops}
\end{center}
\footnotetext{\url{https://projects.propublica.org/politwoops/}}

\begin{center}
\includegraphics[scale=0.5]{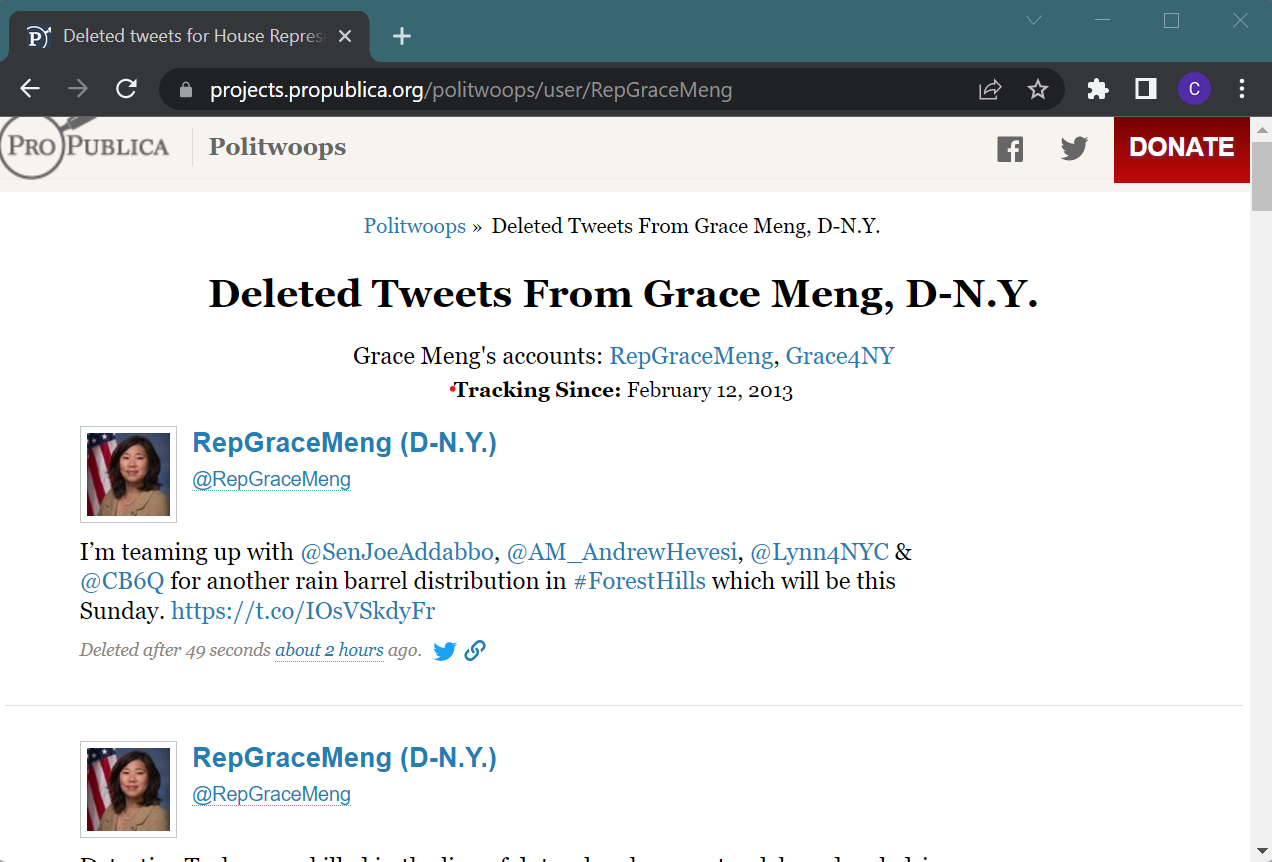}%
\captionof{figure}{The Politwoops page for an elected official whose deleted tweets are being tracked.\protect\footnotemark}\label{labelname}%
\end{center}
\footnotetext{\url{https://projects.propublica.org/politwoops/user/repgracemeng}}

\section{Related Work}
Much work has been done in the field of establishing authorship attribution of social media posts, as well as in the detection of disinformation/misinformation present in a screenshot. This research proposes a novel approach of combining these topics in an effort to detect misattribution disinformation on social media specifically. 

Theophilo et al. \cite{theophilo2021authorship} describe work done in an effort to establish authorship attribution of ``tiny messages," or more specifically, individual tweets. This work involved building deep learning models designed to identify the author of a given text/message using solely that message and no other associated metadata. Methods were developed both to handle the scenario of attributing an author based on only one post, as well as attributing the author of several posts with the assumption that they were all made by the same author. This work differs from this project, in that the original input to their system is the text of a tweet that was definitely made, while our input would be a screenshot of a social media post that could potentially be fabricated. This work focuses on establishing the authenticity of a social media screenshot, while Theophilo et al. focused on establishing the author of a  ``tiny message," that is known to be authentic.

Rocha et al. \cite{7555393} describe similar work done on establishing authorship attribution for the purpose of forensic investigations. Their methodology depended on the deployment of stylometry, that being the analysis of stylistic features of a text in order to identify its author. This involved deploying machine learning techniques to develop a system capable of determining whether or not the author of a message can be identified among a list of known authors based solely on the stylistic choices of their writing. This work is similar to that of Theophilo et al., as their focus was also on establishing specific authorship attribution, as opposed to establishing authenticity of a supposed social media post. 

As far as the field of detecting disinformation in screenshots goes, Abdali et al. \cite{abdali2021identifying} have done work in detecting misinformation present in a screenshot of a news article. This research focused on visual cues present in an article screenshot rather than the actual text of the article itself by analyzing domain-level features as opposed to article-level features. They worked to establish fake news articles by analyzing the source of the article instead of the content of it. This involved the development of a system they dubbed \emph{VizFake} that leveraged a semi-supervised tensor-based approach of classifying the veracity of a news article based on a screenshot of that article. While our work also focuses on detecting disinformation from a screenshot, the goal of \emph{SSauth} is to detect misattribution disinformation in a screenshot of a social media post and the goal of \emph{VizFake} is to detect fake news from a screenshot of a news article.

Using a combination of the live Web and archived Web, Nelson \cite{russellwestbrookblog} detailed work done in an effort to manually verify the authenticity of tweets made after a verbal altercation between an Oklahoma City basketball player (Russell Westbrook) and a Utah Jazz fan (Shane Keisel). This was accomplished by using techniques in web archiving forensics. In summary, Keisel's actual Twitter account was identified shortly after the altercation. This quickly led to Keisel protecting his account. The account was deleted entirely shortly after. However, a series of fake accounts were created meant to impersonate Keisel, with one being particularly effective since it used a Twitter handle that was practically indistinguishable from Keisel's real account. This account made a series of offensive tweets in Keisel's name before eventually being deleted. While extensive research into archived pages of these accounts cleared Keisel's name, most users who engaged with these fake tweets fully believed that he made those offensive remarks. The automation of these types of techniques is a specific goal of this research.

\section{Methodology}
Our methodology for \emph{SSAuth} so far has focused on the automation of querying the Web for evidence of a given tweet, as well as parsing that evidence to determine whether or not it supports or refutes attribution of that tweet. This involved finding sites that we can accept evidence from, refining ways to query for that evidence, and developing software to automate those queries and parse any results. This software is publically available on Github.\footnote{\url{https://github.com/oduwsdl/SSAuth}}

\subsection{Modular design of \emph{SSAuth}}
\emph{SSAuth} needed to automate the process of making search queries to find evidence of a tweet and subsequently evaluating any found evidence for a truth rating. These requirements motivated a modular design, that being a design that allows for the development of modules that handle making queries to different services, as well as modules that handle the scraping of articles from specific sites found from those queries. This type of design makes the addition of compatibility with other search services and sites a simple process of adding more modules to the system in order to handle them. These modules will run independently of each other and will function with the same appropriate inputs. The work done on the modules done so far is detailed below. 

\subsubsection{Work on Snopes and Reuters modules}
Due to the similarity of both their built-in search functions and the structure of their articles, the fundamental algorithms developed for both Snopes and Reuters to query their search engines and parse their fact-check articles are very similar. Queries to find articles about the same fabricated Donald Trump are successful on both sites (Figures \ref{fig:reutersmanquery}, \ref{fig:snopesmanquery}). We developed Python scripts for both sites that automatically construct and make queries within their respective length restrictions given only the body of a tweet and retrieves links to all results. From there, these tools will also scrape those results for a truth rating. (Listings \ref{lst:reuters}, \ref{lst:snopes})\\

\begin{center}
\includegraphics[scale=0.5]{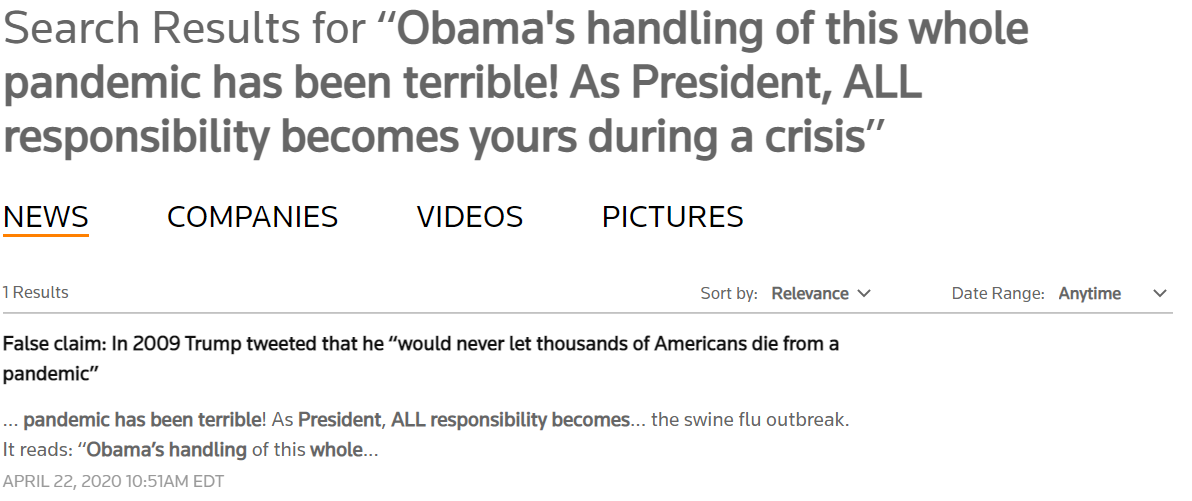}%
\captionof{figure}{A query made to Reuters to find an article about a Donald Trump tweet.\protect\footnotemark}
\label{fig:reutersmanquery}
\end{center}
\footnotetext{\url{https://www.reuters.com/search/news?sortBy=&dateRange=&blob=Obama\%27s++handling+of+this+whole+pandemic+has+been+terrible\%21+As+President\%2C+ALL++responsibility+becomes+yours+during+a+crisis+like}}

\begin{center}
\begin{lstlisting}[caption={Same query as above but, done programatically via our Python script, which is passed the text of a tweet via a command line argument}, label={lst:reuters}]
cs_cbrad022@callisto:~/ToolsForReuters$ python3 query_reuters.py "Obama's 
handling of this whole pandemic has been terrible! As President, ALL 
responsibility becomes yours during a crisis like this, whether or not 
you're entirely to blame. John McCain, and for that matter myself, 
would never let thousands of Americans die from a pandemic while in 
office."
Article found at URL: http://reuters.com/article/idUSKCN2242AK
Truth rating: False
\end{lstlisting}
\end{center}

\begin{center}
\includegraphics[scale=0.5]{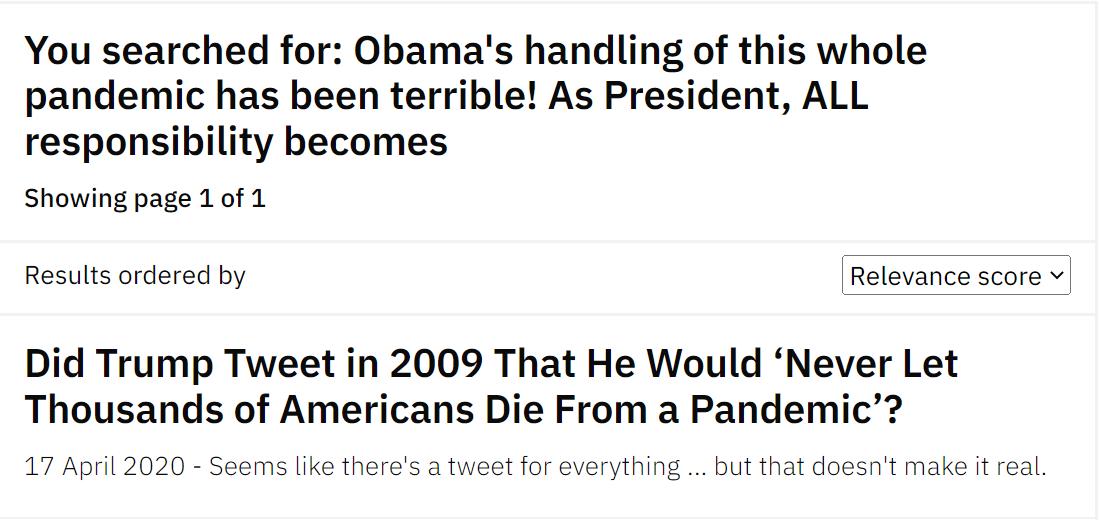}%
\captionof{figure}{A query made to Snopes to find an article about a Donald Trump tweet.\protect\footnotemark }
\label{fig:snopesmanquery}
\end{center}
\footnotetext{\url{https://www.snopes.com/search/Obama's\%20\%20handling\%20of\%20this\%20whole\%20pandemic\%20has\%20been\%20terrible!\%20As\%20President\%2C\%20ALL\%20\%20responsibility\%20becomes/}}

\begin{center}
\begin{lstlisting}[caption={Same query as above, but done programatically via our Python script, which is passed the text of a tweet via a command line argument}, label={lst:snopes}]
cs_cbrad022@europa:~/ToolsForSnopes$ python3 query_snopes.py "Obama's
handling of this whole pandemic has been terrible! As President, ALL 
responsibility becomes yours during a crisis like this, whether or not 
you're entirely to blame. John McCain, and for that matter myself, 
would never let thousands of Americans die from a pandemic while in 
office."
Article found at URL: https://www.snopes.com/fact-check/2009-trump-
tweet-pandemic/
Truth rating: False
\end{lstlisting}
\end{center}

\subsubsection{Work on Politwoops module}
In order to leverage the growing collection of deleted tweets by political officials being stored on Politwoops, we developed software to automatically make search queries to the Politwoops built-in search engine in an effort to find whether or not a specific tweet is present on the site. This included extensive experimentation \cite{politwoopsblog} to find the most efficient queries to find a tweet that is known to be stored there. We found that queries using only the first 50 characters of a tweet showed efficacy (Figure \ref{fig:politwoopsmanquery}), but other potentially more effective methods exist and future work will entail exploring those methods further. Currently our software constructs queries using the first 50 characters of a given tweet and compares any results of the query against any results found (Listing  \ref{lst:politwoops}).

\begin{center}
\includegraphics[scale=0.5]{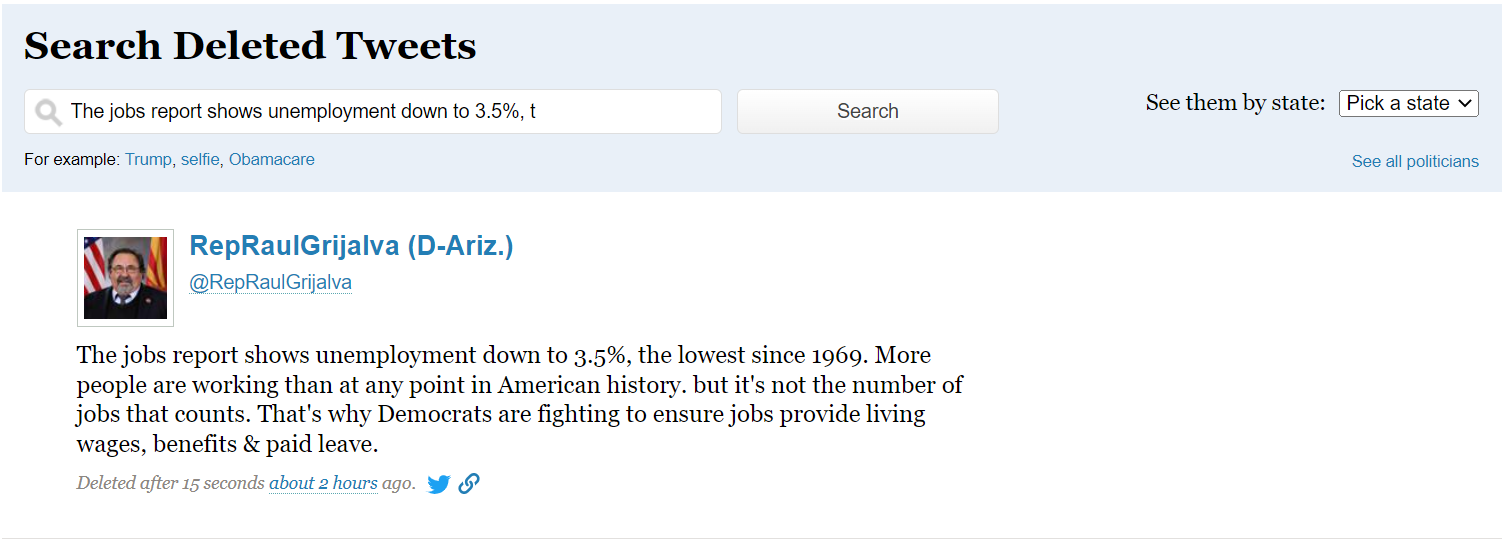}%
\captionof{figure}{A successful query to Politwoops using first 50 characters of a tweet.\protect\footnotemark}
\label{fig:politwoopsmanquery}
\end{center}
\footnotetext{\url{https://projects.propublica.org/politwoops/index?utf8=\%E2\%9C\%93&q=The+jobs+report+shows+unemployment+down+to+3.5\%25\%2C+t}}

\begin{center}
\begin{lstlisting}[caption={Same query as above, but done programatically via a Python script, which is passed the text of a tweet to be searched for via a command line argument.}, label={lst:politwoops}]
cs_cbrad022@callisto:~/ToolsForPolitwoops$ python3 query_politwoops.py 
"The jobs report shows unemployment down to 3.5%, the lowest since 1969. 
More people are working than at any point in American history. but it's 
not the number of jobs that counts. That's why Democrats are fighting to 
ensure jobs provide living wages, benefits & paid leave."
That tweet was successfully queried on Politwoops
\end{lstlisting}
\label{fig:politwoopsautoquery}
\end{center}

\subsubsection{Work on Google module}
Our research also led to the development of software that is capable of querying Google and scraping the SERP for links to relevant results. This software utilizes a library developed by Doctor Alexander Nwala during his time at Old Dominion University \cite{nwala-scraper}. Our Python script is capable of taking in a supposed tweet as a command line argument, making a valid query, and scraping any pages found that we are currently able to parse for a truth rating (Listing \ref{lst:google}).
\begin{center}
\begin{lstlisting}[caption={An automated Google query that is passed the text of a tweet via a command line argument.}, label={lst:google}]
cs_cbrad022@europa:~/ToolsForGoogle$ python3 query_google.py "Obama's 
handling of this whole pandemic has been terrible! As President, ALL 
responsibility becomes yours during a crisis like this, whether or not 
you're entirely to blame. John McCain, and for that matter myself, 
would never let thousands of Americans die from a pandemic while in 
office."

Reuters article found at URL: https://www.reuters.com/article/uk-fact
check-trump-tweet-thousands-die/false-claim-in-2009-trump-tweeted-that
-he-would-never-let-thousands-of-americans-die-from-a-pandemic-idUSKCN
2242AK
Verdict found in this article is False

Snopes article found at URL: https://www.snopes.com/fact-check/2009-tr
ump-tweet-pandemic/
Truth rating in this article is False
\end{lstlisting}
\end{center}

\subsection{What we can currently uncover}
When used in combination, these tools can uncover information about the existence of evidence of a supposed tweet. When evidence is found, we can likely come to a reasonable conclusion about the veracity of that tweet. The use of this evidence will ultimately be used to make a determination of either true (the tweet was verifiably made) (Figure \ref{fig:unpresidented}) or false (we can prove the tweet was not made) (Figure \ref{fig:trumpscreenshot}). 

\begin{center}
\includegraphics[scale=0.5]{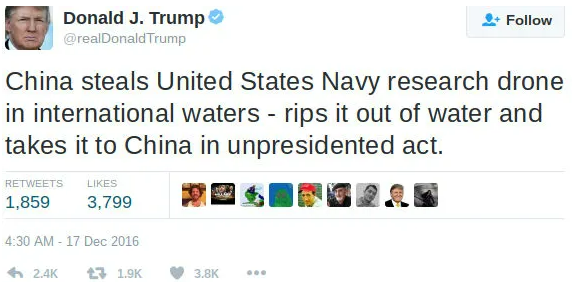}%
\captionof{figure}{A screenshot of a real deleted Donald Trump tweet. Evidence of its existence is found on Snopes and Politwoops.\protect\footnotemark}
\label{fig:unpresidented}
\end{center}
\footnotetext{\url{https://www.snopes.com/fact-check/trump-sends-unpresidented-tweet/}, \\ \url{https://projects.propublica.org/politwoops/tweet/810099766063493120}}

\begin{center}
\includegraphics[scale=0.5]{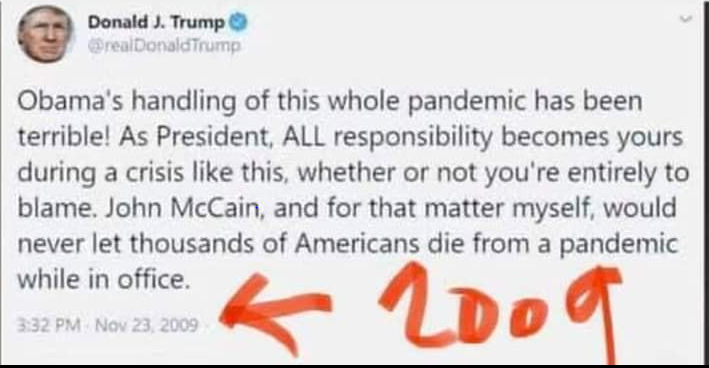}%
\captionof{figure}{A screenshot of the fake Donald Trump tweet previously mentioned. Evidence is found proving fabrication on Snopes and Reuters.\protect\footnotemark}
\label{fig:trumpscreenshot}
\end{center}
\footnotetext{\url{https://www.snopes.com/fact-check/2009-trump-tweet-pandemic/}, \\ \url{https://www.reuters.com/article/idUSKCN2242AK}}

Another possibility is that we do not find any evidence, meaning that we cannot make any conclusion about the existence of the tweet, regardless of whether or not the tweet is real (Figure \ref{fig:trumpreal}) or fabricated (Figure \ref{fig:mytrumptweet}). 

\begin{center}
\includegraphics[scale=0.5]{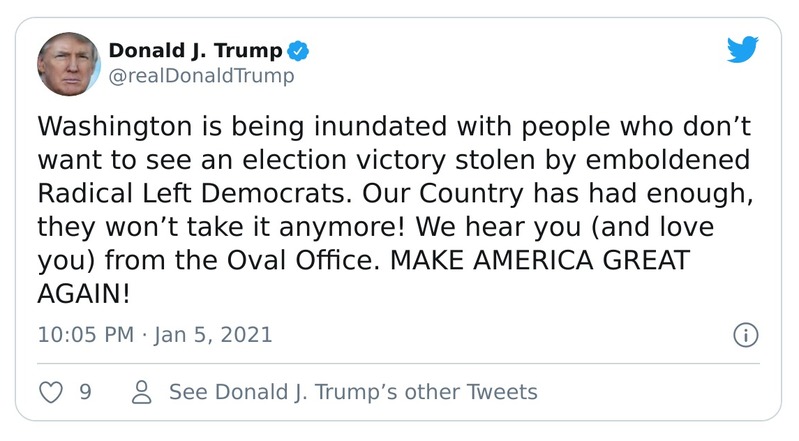}%
\captionof{figure}{A screenshot of an authentic Donald Trump tweet that cannot currently be verified using our tools}
\label{fig:trumpreal}
\end{center}

\begin{center}
\includegraphics[scale=0.5]{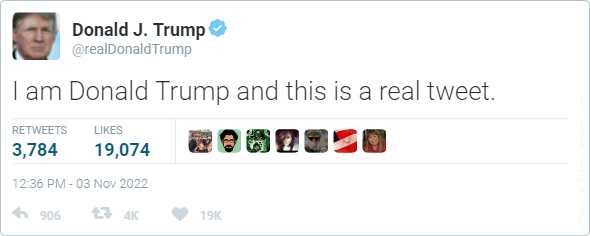}%
\captionof{figure}{A screenshot of a fabricated tweet that cannot currently be verified using our tools}
\label{fig:mytrumptweet}
\end{center}

\section{Evaluation of current \emph{SSAuth} modules}
The variability of the ability of search engines to find the evidence we are looking for motivated the evaluation of the search engines we are using by building queries to search for evidence that we know exists to see if those search engines are effective at finding that evidence.
\subsection{Building a ground truth dataset of tweets with known evidence online}
Evaluation of these search engines required us to begin construction of a  ground truth dataset of tweets with known evidence on the Web that we should theoretically be able to search for. So far, the dataset consists of 30 tweets that have Snopes.com fact-check articles relating to those tweets. These fact-check articles include both articles concerning attribution of the tweet in question, as well as articles concerning the actual validity of a statement made in a tweet. In addition to the bodies of those tweets, the dataset contains the URLs for their respective Snopes articles as well as an indication of whether or not the tweet is authentic according to the article. Entries for all authentic tweets also contain archived and live URLs for the tweets themselves. The dataset currently contains 15 fabricated tweets and 15 authentic tweets. This dataset can and should be added to in the future with more tweets with known evidence online. The dataset is available on Github\footnote{\url{https://github.com/oduwsdl/SSAuth}}

\subsection{Evaluation of search engines on the dataset}
The evaluation of queries made to find the articles in this dataset was completed by manually making valid queries using each search engine being evaluated. Based on the SERPs of those queries, $MRR$ and $P@1$ (Figures \ref{fig:tedcruzmetric}, \ref{fig:giulianimetric}, \ref{fig:beyoncemetric}) values were computed and logged. Queries made to the Reuters search engine are included as well in order to see the overlap between Snopes and Reuters fact-check articles. However, these metrics are subject to change due to differences in SERPs requested over time. However, due to the ever-changing nature of the Web, these metrics are subject to change as requested SERPs change overtime. The search engines being evaluated include Google, Google with the site:snopes.com operator(filters out results that are not from Snopes.com), the Snopes built-in search engine, and the Reuters built-in search engine. The Reuters search engine is included here in order to find if any tweets had related fact-check articles on both Snopes and Reuters.

\begin{center}
\includegraphics[scale=0.5]{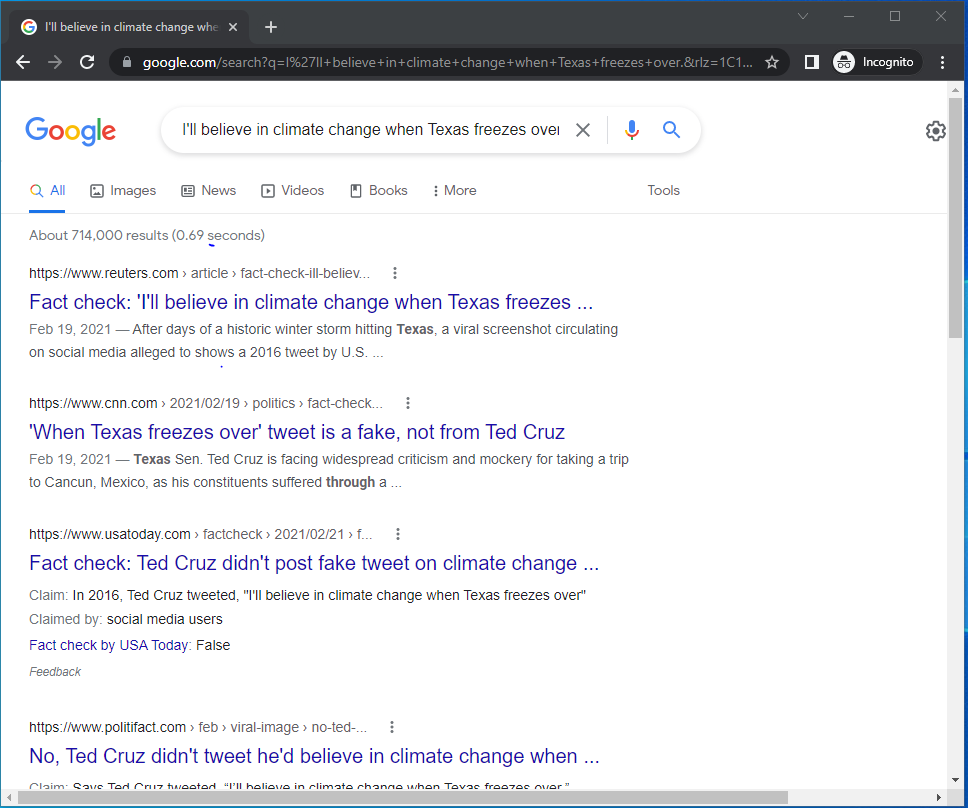}%
\captionof{figure}{A Google query for a tweet with $P@1=1$ and $MRR=1$}
\label{fig:tedcruzmetric}
\end{center}

\begin{center}
\includegraphics[scale=0.5]{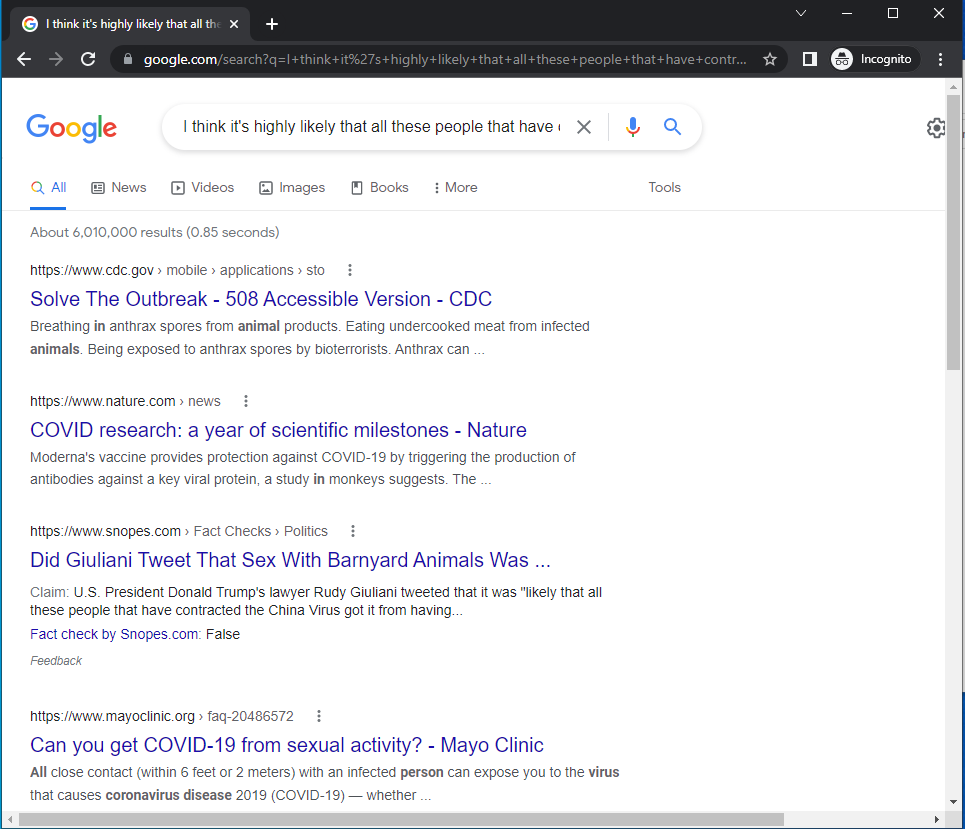}%
\captionof{figure}{A Google query for a tweet with a $P@1=0$ and $MRR=0.33$}
\label{fig:giulianimetric}
\end{center}

\begin{center}
\includegraphics[scale=0.5]{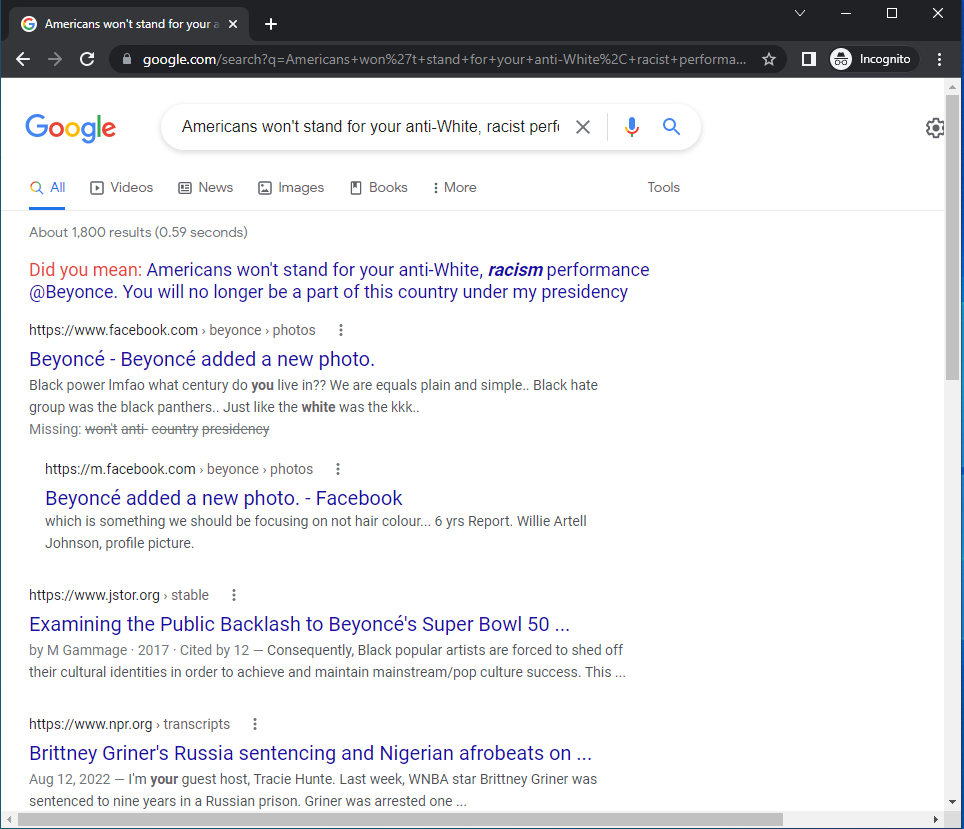}%
\captionof{figure}{A Google query for a tweet with $P@1=0$ and $MRR=0$}
\label{fig:beyoncemetric}
\end{center}

\begin{table}[ht]
\centering
\begin{tabular}{|l|l|l|}
\hline
Search engine being evaluated & $MRR$ & Mean $P@1$ \\ \hline
Google                        & 0.4513               & 0.2667                 \\ \hline
Google with site:snopes.com   & 0.8667               & 0.8667                 \\ \hline
Snopes built-in search        & 0.5500               & 0.5333                 \\ \hline
Reuters built-in search       & 0.2167               & 0.2000                 \\ \hline
\end{tabular}
\caption{Results of analysis of queries done to search for current dataset of 30 Snopes articles relating to tweets}
\end{table}

\section{Future work}
Building off of this research will entail adding more modules to the design, including one that can derive a truth rating from any generalized fact-check article. The modules also need to be connected into a cohesive system that executes all queries in the system and parses all of the results given the body of a supposed tweet. More work needs to be done on the dataset we began constructing by adding more tweets with known evidence on the Web. Software should also be developed that will automate the evaluation of our software on the dataset. 

We have initially focused on tweet attribution, but the project as a whole aims to verify attribution for a screenshot of any social media post. For this reason, methodology for verifying screenshots of social media posts from other sites such as Facebook and Instagram need to be developed as well. 

In the future, this work will also entail the usage of image processing techniques to classify the site that a screenshot of a social media post is taken from, and extracting the relevant metadata from that screenshot accordingly. 

\section{Conclusions}
Misattribution disinformation is a unique form of disinformation that has become more prevalent as the practice of social media users sharing screenshots of posts has become a necessity online. In order to combat this type of disinformation, this project aims to develop \emph{SSAuth}, a system that is capable of taking in a screenshot of a social media post as input and determine whether or not the post depicted in the screenshot is correctly attributed or not. Our research so far has focused on the development of methods to query the Web for evidence of correct attribution of a tweet given only the body of that tweet. We developed software to query various search services that contain relevant information, including the built-in search engine for Snopes, Reuters, and Politwoops. We developed tools to query Google as well. In order to evaluate the efficacy of these search engines in finding evidence that we know to exist, we began construction of a dataset of tweets that have known evidence on the Web. Our collection of metrics based on queries made to find the Snopes articles corresponding to tweets in the current ground truth dataset showed that the Snopes built-in search engine worked yielded $MRR=0.5500$ and $P@1=0.5333$. Our metrics also showed that Google queries filtered with the site:snopes.com operator were generally effective at finding the Snopes articles with $MRR=0.8667$ and $P@1=0.8667$. All of the tools developed, as well as the current ground truth dataset is available publicly on Github.\footnote{\url{https://github.com/oduwsdl/SSAuth}}

\section{Acknowledgements}
This work was supported in part through the National Science Foundation Computer and Information Science and Engineering Research Experience for Undergraduates Site Award \#2149607.\footnote{\url{https://oducsreu.github.io/}} 

Our research received generous assistance from several graduate students throughout the course of the summer program, namely Emily Escamilla and Tarannum Zaki, who are both PhD students at ODU. Much of our work would also not have been possible without the software developed by Dr. Alexander Nwala. Lastly, we would also like to acknowledge the assistance given by fellow members of the 2022 REU cohort \cite{odu-reu-2022-1, odu-reu-2022-2}.

\newpage
\renewcommand{\refname}{References}
\bibliographystyle{ieeetr}
\bibliography{refs}

\begin{thebibliography}{10}

\bibitem{twitter-suspends-trump}
{Twitter}, ``Permanent suspension of {@realDonaldTrump}.''
  \url{https://blog.twitter.com/en_us/topics/company/2020/suspension}, 2021.

\bibitem{garg2021replaying}
K.~Garg, H.~R. Jayanetti, S.~Alam, M.~C. Weigle, and M.~L. Nelson, ``Replaying
  archived {Twitter}: When your bird is broken, will it bring you down?,'' in
  {\em 2021 ACM/IEEE Joint Conference on Digital Libraries (JCDL)},
  pp.~160--169, IEEE, 2021.

\bibitem{kriesberg2022second}
A.~Kriesberg and A.~Acker, ``The second {US} presidential social media
  transition: How private platforms impact the digital preservation of public
  records,'' {\em Journal of the Association for Information Science and
  Technology}, vol.~73, no.~11, pp.~1529--1542, 2022.

\bibitem{theophilo2021authorship}
A.~Theophilo, R.~Giot, and A.~Rocha, ``Authorship attribution of social media
  messages,'' {\em IEEE Transactions on Computational Social Systems}, 2021.

\bibitem{7555393}
A.~Rocha, W.~J. Scheirer, C.~W. Forstall, T.~Cavalcante, A.~Theophilo, B.~Shen,
  A.~R.~B. Carvalho, and E.~Stamatatos, ``Authorship attribution for social
  media forensics,'' {\em IEEE Transactions on Information Forensics and
  Security}, vol.~12, no.~1, pp.~5--33, 2017.

\bibitem{abdali2021identifying}
S.~Abdali, R.~Gurav, S.~Menon, D.~Fonseca, N.~Entezari, N.~Shah, and E.~E.
  Papalexakis, ``Identifying misinformation from website screenshots.,'' in
  {\em International AAAI Conference on Web and Social Media}, pp.~2--13, 2021.

\bibitem{russellwestbrookblog}
M.~L. Nelson, ``{Russell Westbrook}, {Shane Keisel}, fake twitter accounts, and
  web archives.''
  \url{https://ws-dl.blogspot.com/2019/04/2019-04-17-russell-westbrook-shane.html},
  2019.

\bibitem{politwoopsblog}
C.~Bradford, ``Querying the {Politwoops} search engine for deleted tweets.''
  \url{https://ws-dl.blogspot.com/2022/09/2022-09-15-querying-politwoops-search.html},
  2022.

\bibitem{nwala-scraper}
A.~Nwala, ``scraper.'' \url{https://github.com/oduwsdl/scraper}, 2019.

\bibitem{odu-reu-2022-1}
Y.~Abeysinghe, ``Disinformation detection and analytics {REU} program - mid
  summer presentations.''
  \url{https://ws-dl.blogspot.com/2022/07/2022-07-15-disinformation-detection-and.html},
  2022.

\bibitem{odu-reu-2022-2}
B.~Farrow, ``Disinformation detection and analytics {REU} program - final
  summer presentations.''
  \url{https://ws-dl.blogspot.com/2022/08/2022-08-19-disinformation-detection-and.html},
  2022.

\end{thebibliography}

\newpage

\end{document}